# Constructing Cryptographic Multilinear Maps Using Affine Automorphisms

By Paul Hriljac

The point of this paper is to use affine automorphisms from algebraic geometry to build cryptographic multivariate mappings, see [BS]. We will construct groups $\mathcal{G}, \mathcal{H}$, both isomorphic to the cyclic group $\mathbb{Z}/p\mathbb{Z}$, and multilinear pairings $\mathcal{G}^k \to \mathcal{H}$. The construction is reminiscent of techniques in multivariate encryption ([M1], [MI], [P1]). We display several different versions of the discrete logarithm problem for these groups. We show that the efficient solution of some of these problems result in efficient algorithms for inverting systems of multivariate polynomials corresponding to affine automorphisms, which implies that such problems are as computationally difficult as breaking multivariate encryption. The author would like to thank Dan Boneh for some helpful comments.

The method of constructing these maps is derived from ideas in [Hr]. The gist of the construction of each group will be a collection of vectors and polynomials depending on parameters $p, n, d$. The parameter $p$ will be a prime number and is the order of the group. The parameter $n$ is the dimension of the vector space the group is embedded into as a set. The group will be a proper subset of the vector space, it will be presented by one generating element and a system of multivariate polynomials describing the group operation. The parameter $d$ controls the degree of those polynomials. The key space for these groups will be a set of non-linear polynomial transformations on $n$-vectors with coefficients in GF($p$). The size of this key space grows polynomially with $p$, exponentially with $d$, and doubly exponentially with $n$. For an algebraic circuit $A$, let $f_A$ denote the associated output polynomials and let $\mathcal{E}[A]$ be the equivalent polynomials expressed in sparse form. We will use the operator $\mathcal{E}$ to rewrite compositions of algebraic transformations, so the resulting expression consist of multivariable polynomials written sparsely.

**Definition**: *let $p$ be a prime and let $\mathbb{F} = GF(p)$. Let $n$ be a positive integer and $C$ be a nonsingular affine rational curve in affine $n$-space which is the image of a linear immersion $\rho: \mathbb{A}^1 \to \mathbb{A}^n$. Denote by $\rho^{-1}$ the inverse map $C \to \mathbb{A}^1$. Since $\rho$ is a linear immersion, $\rho^{-1}$ can be represented by a linear map $\rho^{-1}: \mathbb{A}^n \to \mathbb{A}^1$. Let $\theta$ be an automorphism of $\mathbb{A}^n$ and let $D = \theta(C)$ be another affine rational curve which is the image of the first curve by the automorphism. Let $\mathcal{G} = D(\mathbb{F})$ be the rational points of $D$, considered as vectors in $\mathbb{F}^n$. We put a group law on $\mathcal{G}$ by borrowing the group law on $\mathbb{A}^1$: For elements $P, Q \in \mathcal{G}$ let*

$$\mathcal{g}(P,Q) = \mathcal{E}\left[\theta\left(\rho\left(\rho^{-1}(\theta^{-1}(P)) + \rho^{-1}(\theta^{-1}(Q))\right)\right)\right]$$

With this definition we get a group which is isomorphic to $\mathbb{Z}/p\mathbb{Z}$ but is represented by a set of vectors in $\mathbb{F}^n$ and with a group law that is expressed by non-linear algebraic functions of the vector components. Since $\mathcal{G}$ is cyclic choosing any element of $\mathcal{G}$ other than the identity will result in a generator. Note that such a group law can be expressed in the coordinates of the ambient space of the curve $D$:

**Definition**: *For $x, y \in \mathbb{F}^n$ let $\mathcal{g}(x,y) = \mathcal{E}\left[\theta\left(\rho\left(\rho^{-1}(\theta^{-1}(x)) + \rho^{-1}(\theta^{-1}(y))\right)\right)\right].$*

**Proposition:** *If $x, y \in \mathcal{G}$, then $\mathcal{g}(x,y) \in \mathcal{G}$.*

**Note:** Sometimes, for $P, Q \in \mathcal{G}$ we express those elements as points on the associated curve $D$ and write $P + Q$ in place of $\mathcal{g}(P, Q)$.

**Note:** These groups can either be specified with a generator $P_0$ and a group law $\mathcal{g}(x, y)$ or as linear immersion $\rho: \mathbb{A}^1 \to \mathbb{A}^n$ and an automorphism $\theta$ of $\mathbb{A}^n$.

**Definition**: *Given a linear embedding $\rho: \mathbb{A}^1 \to \mathbb{A}^n$ and an automorphism $\theta: \mathbb{A}^n \tilde{\to} \mathbb{A}^n$, both over the field $\mathbb{F}$ = GF(p), define a cyclic group $G$ of order $p$ by a generator $P_0 \in \mathbb{F}^n$ and a group law $g$ given by a system of polynomials as above. Define $\mathfrak{G}_{p,d,n}$ as the set of group structures using automorphisms $\theta: \mathbb{A}^n \tilde{\to} \mathbb{A}^n$ over GF(p) with the total degree of each component polynomial $\leq d$. Let $\mathfrak{G}_{p,n} = \bigcup_d \mathfrak{G}_{p,d,n}$ and $\mathfrak{G}_p = \bigcup_n \mathfrak{G}_{p,n}$.*

**Example 1**: Let $\mathbb{F} = GF(1801)$ and $n = 3$. Let

$$\rho: t \to \begin{bmatrix} t \\ t-1 \\ t+1 \end{bmatrix}$$

Then $C$ is an affine line in 3-space, defined by the relations $x_2 = x_1 - 1$, $x_3 = x_1 + 1$. For points $\begin{bmatrix} x_1 \\ x_2 \\ x_3 \end{bmatrix} \in C$ the mapping $\rho^{-1}$ is just $\rho^{-1}\left(\begin{bmatrix} x_1 \\ x_2 \\ x_3 \end{bmatrix}\right) = x_1$. We create an automorphism by composing three simple automorphisms together, two are just linear transformations:

$$\lambda_1(x_1, x_2, x_3) = \begin{bmatrix} 1 & 0 & 1 \\ 1 & 1 & 0 \\ 0 & 1 & 0 \end{bmatrix} \begin{bmatrix} x_1 \\ x_2 \\ x_3 \end{bmatrix}$$

$$\lambda_2(x_1, x_2, x_3) = \begin{bmatrix} 0 & 1 & 1 \\ 1 & 0 & 1 \\ 0 & 1 & 0 \end{bmatrix} \begin{bmatrix} x_1 \\ x_2 \\ x_3 \end{bmatrix}$$

The third is a triangular automorphism $\tau\left(\begin{bmatrix} x_1 \\ x_2 \\ x_3 \end{bmatrix}\right) = \begin{bmatrix} x_1 + 1 \\ x_2 + x_1 \\ x_3 + x_1 x_2 + x_1^2 \end{bmatrix}$. Let $\theta = \lambda_1 \circ \tau \circ \lambda_2$, $D = \theta(C)$ and let $G$ be the rational points of $D$. Then $G$ is a set of 1801 $\mathbb{F}$-vectors of length 3, one of which is $\begin{bmatrix} 7 \\ 7 \\ 4 \end{bmatrix}$, which can be taken as a generator for the group.

The group law on $G$ is then given by $g\left(\begin{bmatrix} x_1 \\ x_2 \\ x_3 \end{bmatrix}, \begin{bmatrix} y_1 \\ y_2 \\ y_3 \end{bmatrix}\right) = \begin{bmatrix} z_1 \\ z_2 \\ z_3 \end{bmatrix}$ with $z_1 =$

$1799 + 1791\,x_1 + 5\,x_2 + 10\,x_3 + 32\,x_1 x_2 y_1 y_2 + 1769\,x_1 y_3 + 16\,x_1 y_1^2 + 16\,x_1 y_2^2 + 1785\,x_2 y_1 + 8\,x_2 y_2$
$+ 16\,x_2 y_3 + 1793\,x_2 y_1^2 + 1793\,x_2 y_2^2 + 1769\,x_3 y_1 + 16\,x_3 y_2 + 32\,x_3 y_3 + 1785\,x_3 y_1^2 + 1785\,x_3 y_2^2$
$+ 16\,x_1^2 y_1 + 1793\,x_1^2 y_2 + 1785\,x_1^2 y_3 + 8\,x_1^2 y_1^2 + 8\,x_1^2 y_2^2 + 16\,x_2^2 y_1 + 1793\,x_2^2 y_2 + 1785\,x_2^2 y_3 + 24\,x_1^2 x_2^2 + 1785\,x_1^3 x_2 + 1785\,x_1 x_2^3 + 1793\,x_1^3 x_3 + 1777\,y_2^2 y_3 + 1785\,y_2 y_3^2 + 32\,x_1 y_1 + 1785\,x_1 y_2$
$+ 1774\,x_1 x_3 + 11\,x_2 x_3 + 1795\,y_1 y_2 + 16\,x_1 x_2 y_1 y_3 + 1785\,x_1 x_2 y_2 y_3 + 16\,x_1 x_3 y_1 y_2 + 8\,x_1 x_3 y_1 y_3$
$+ 1793\,x_1 x_3 y_2 y_3 + 1785\,x_2 x_3 y_1 y_2 + 1793\,x_2 x_3 y_1 y_3 + 8\,x_2 x_3 y_2 y_3 + 1795\,x_1 x_2 + 56\,y_1 y_2 y_3$
$+ 24\,x_1^2 x_2 x_3 + 1777\,x_1 x_2^2 x_3 + 1793\,x_1 x_2 x_3^2 + 1785\,x_3 y_2 y_3 + 1769\,x_1 x_2 y_1 + 1785\,x_2 x_3 y_3$
$+ 16\,x_1 x_2 y_2 + 32\,x_3 y_1 y_2 + 1785\,x_1 x_2 y_1^2 + 1785\,x_1^2 y_1 y_2 + 1793\,x_2 y_2 y_3 + 1769\,x_1 y_1 y_2 + 8\,x_1 x_3 y_2$
$+ 1793\,x_2 x_3 y_2 + 1785\,x_1 y_1 y_3 + 16\,x_2 y_1 y_2 + 16\,x_1 y_2 y_3 + 1793\,x_1 x_3 y_2^2 + 1785\,x_1 x_3 y_1$
$+ 1793\,x_1 x_3 y_1^2 + 16\,x_1 x_3 y_3 + 8\,x_2 y_1 y_3 + 8\,x_1^2 y_2 y_3 + 1793\,x_2^2 y_1 y_3 + 32\,x_1 x_2 y_3 + 16\,x_2 x_3 y_1$
$+ 1793\,x_1^2 y_1 y_3 + 16\,x_3 y_1 y_3 + 8\,x_2^2 y_2 y_3 + 1785\,x_1 x_2 y_2^2 + 1785\,x_2^2 y_1 y_2 + 8\,x_2 x_3 y_1^2 + 8\,x_2 x_3 y_2^2$
$+ 24\,y_1^2 y_2 y_3 + 1777\,y_1 y_2^2 y_3 + 1793\,y_1 y_2 y_3^2 + 56\,x_1 x_2 x_3 + 16\,x_3^2 + 11\,x_1^2 + 1791\,y_1 + 5\,y_2 + 10\,y_3$
$+ 11\,y_1^2 + 1800\,x_2^2 + 16\,y_1^3 + 4\,x_1^4 + 4\,x_2^4 + 1800\,y_2^2 + 16\,y_3^3 + 1793\,y_2^3 + 4\,y_1^4 + 4\,y_2^4 + 16\,x_1^3 + 1793\,x_2^3 + 4\,x_1^2 x_3^2 + 8\,x_2^3 x_3 + 4\,x_2^2 x_3^2 + 1761\,y_1^2 y_2 + 1769\,y_1^2 y_3 + 32\,y_1 y_2^2 + 16\,y_1 y_3^2 + 1774\,y_1 y_3$
$+ 11\,y_2 y_3 + 8\,x_2^2 y_1^2 + 8\,x_2^2 y_2^2 + 1785\,y_1^3 y_2 + 1793\,y_1^3 y_3 + 24\,y_1^2 y_2^2 + 4\,y_1^2 y_3^2 + 1785\,y_1 y_2^3 + 8\,y_2^3 y_3$
$+ 4\,y_2^2 y_3^2 + 1761\,x_1^2 x_2 + 1769\,x_1^2 x_3 + 32\,x_1 x_2^2 + 16\,x_1 x_3^2 + 1777\,x_2^2 x_3 + 1785\,x_2 x_3^2$

$z_2 =$

$1795\,x_1 + 3\,x_2 + 6\,x_3 + 32\,x_1 x_2 y_1 y_2 + 1769\,x_1 y_3 + 16\,x_1 y_1^2 + 16\,x_1 y_2^2 + 1785\,x_2 y_1 + 8\,x_2 y_2 + 16\,x_2 y_3$
$+ 1793\,x_2 y_1^2 + 1793\,x_2 y_2^2 + 1769\,x_3 y_1 + 16\,x_3 y_2 + 32\,x_3 y_3 + 1785\,x_3 y_1^2 + 1785\,x_3 y_2^2 + 16\,x_1^2 y_1$
$+ 1793\,x_1^2 y_2 + 1785\,x_1^2 y_3 + 8\,x_1^2 y_1^2 + 8\,x_1^2 y_2^2 + 16\,x_2^2 y_1 + 1793\,x_2^2 y_2 + 1785\,x_2^2 y_3 + 24\,x_1^2 x_2^2$
$+ 1785\,x_1^3 x_2 + 1785\,x_1 x_2^3 + 1793\,x_1^3 x_3 + 1777\,y_2^2 y_3 + 1785\,y_2 y_3^2 + 32\,x_1 y_1 + 1785\,x_1 y_2$
$+ 1772\,x_1 x_3 + 13\,x_2 x_3 + 1791\,y_1 y_2 + 16\,x_1 x_2 y_1 y_3 + 1785\,x_1 x_2 y_2 y_3 + 16\,x_1 x_3 y_1 y_2 + 8\,x_1 x_3 y_1 y_3$
$+ 1793\,x_1 x_3 y_2 y_3 + 1785\,x_2 x_3 y_1 y_2 + 1793\,x_2 x_3 y_1 y_3 + 8\,x_2 x_3 y_2 y_3 + 1791\,x_1 x_2 + 56\,y_1 y_2 y_3$
$+ 24\,x_1^2 x_2 x_3 + 1777\,x_1 x_2^2 x_3 + 1793\,x_1 x_2 x_3^2 + 1785\,x_3 y_2 y_3 + 1769\,x_1 x_2 y_1 + 1785\,x_2 x_3 y_3$
$+ 16\,x_1 x_2 y_2 + 32\,x_3 y_1 y_2 + 1785\,x_1 x_2 y_1^2 + 1785\,x_1^2 y_1 y_2 + 1793\,x_2 y_2 y_3 + 1769\,x_1 y_1 y_2 + 8\,x_1 x_3 y_2$
$+ 1793\,x_2 x_3 y_2 + 1785\,x_1 y_1 y_3 + 16\,x_2 y_1 y_2 + 16\,x_1 y_2 y_3 + 1793\,x_1 x_3 y_2^2 + 1785\,x_1 x_3 y_1$
$+ 1793\,x_1 x_3 y_1^2 + 16\,x_1 x_3 y_3 + 8\,x_2 y_1 y_3 + 8\,x_1^2 y_2 y_3 + 1793\,x_2^2 y_1 y_3 + 32\,x_1 x_2 y_3 + 16\,x_2 x_3 y_1$
$+ 1793\,x_1^2 y_1 y_3 + 16\,x_3 y_1 y_3 + 8\,x_2^2 y_2 y_3 + 1785\,x_1 x_2 y_2^2 + 1785\,x_2^2 y_1 y_2 + 8\,x_2 x_3 y_1^2 + 8\,x_2 x_3 y_2^2$
$+ 24\,y_1^2 y_2 y_3 + 1777\,y_1 y_2^2 y_3 + 1793\,y_1 y_2 y_3^2 + 56\,x_1 x_2 x_3 + 16\,x_3^2 + 13\,x_1^2 + 1795\,y_1 + 3\,y_2 + 6\,y_3$
$+ 13\,y_1^2 + x_2^2 + 16\,y_1^3 + 4\,x_1^4 + 4\,x_2^4 + y_2^2 + 16\,y_3^3 + 1793\,y_2^3 + 4\,y_1^4 + 4\,y_2^4 + 16\,x_1^3 + 1793\,x_2^3 + 4\,x_1^2 x_3^2$
$+ 8\,x_2^3 x_3 + 4\,x_2^2 x_3^2 + 1761\,y_1^2 y_2 + 1769\,y_1^2 y_3 + 32\,y_1 y_2^2 + 16\,y_1 y_3^2 + 1772\,y_1 y_3 + 13\,y_2 y_3 + 8\,x_2^2 y_1^2$
$+ 8\,x_2^2 y_2^2 + 1785\,y_1^3 y_2 + 1793\,y_1^3 y_3 + 24\,y_1^2 y_2^2 + 4\,y_1^2 y_3^2 + 1785\,y_1 y_2^3 + 8\,y_2^3 y_3 + 4\,y_2^2 y_3^2 + 1761\,x_1^2 x_2$
$+ 1769\,x_1^2 x_3 + 32\,x_1 x_2^2 + 16\,x_1 x_3^2 + 1777\,x_2^2 x_3 + 1785\,x_2 x_3^2$

$z_3 =$

$$1793\, y_1 + 4\, y_2 + 8\, y_3 + 8\, y_1 y_2 + 1797\, y_1^2 + 4\, y_1 y_3 + 1797\, y_2^2 + 1797\, y_2 y_3 + 1793\, x_1 + 4\, x_2 + 8\, x_3$$
$$+ 8\, x_1 x_2 + 1797\, x_1^2 + 4\, x_1 x_3 + 1797\, x_2^2 + 1797\, x_2 x_3$$

Here all operations take place in $\mathbb{F}$ = GF(1801).

Let $m: \mathbb{F}^k \to \mathbb{F}$ be the multilinear form $m(x_1, \ldots, x_k) = \prod_i x_i$, so $m(x_1, \ldots x_i + y_i, \ldots, x_k) = m(x_1, \ldots x_i, \ldots, x_k) + m(x_1, \ldots, y_i, \ldots, x_k)$, for all $i$. Let $\mathcal{G}, \mathcal{H} \in \mathbb{G}_{p,n}$ with associated data $\{\rho, \theta\}, \{\kappa, \sigma\}$

**Define** $\mu: \mathcal{G}^k \to \mathcal{H}$ by $\mu(P_1, \ldots, P_k) = \sigma\left(\kappa\left(m\left(\rho^{-1}(\theta^{-1}(P_1)), \ldots, \rho^{-1}(\theta^{-1}(P_k))\right)\right)\right)$.

*Here we identify elements of $\mathbb{A}^1(\mathbb{F})$ with elements of $\mathbb{F}$.*

**Theorem**: $\mu: \mathcal{G}^k \to \mathcal{H}$ is a multilinear form.

Proof: Let $P_i \in \mathcal{G}$. Let $t_i \in \mathbb{F}$ be the parameter value corresponding to $\rho^{-1}(\theta^{-1}(P_i)) \in \mathbb{A}^1$. Suppose $Q_i$ is another element of $\mathcal{G}$, with corresponding parameter value $u_i \in \mathbb{F}$. Then
$$\mu(P_1, \ldots, P_i + Q_i, \ldots, P_k) =$$
$$\sigma\left(\kappa\big(m(t_1, \ldots, t_i + u_i, \ldots t_k)\big)\right) =$$
$$\sigma\left(\kappa\big(m(t_1, \ldots, t_i, \ldots t_k) + m(t_1, \ldots, u_i, \ldots t_k)\big)\right) =$$
$$\sigma\left(\kappa\big(m(t_1, \ldots, t_i, \ldots t_k)\big)\right) + \sigma\left(\kappa\big(m(t_1, \ldots, u_i, \ldots t_k)\big)\right) =$$
$$\mu(P_1, \ldots, P_i, \ldots, P_k) + \mu(P_1, \ldots, Q_i, \ldots, P_k).$$
QED.

Now starting with two groups $\mathcal{G} \leftrightarrow \{\rho, \theta\}$, and $\mathcal{H} \leftrightarrow \{\kappa, \rho\}$ and the multilinear form $m: \mathbb{F}^k \to \mathbb{F}$ one can rewrite the function $\mu$ in terms of the ambient coordinates the affine rational curves $D, E \subset \mathbb{A}^n$ and obtain a system of polynomial functions $\mu: (\mathbb{A}^n)^k \to \mathbb{A}^n$,
$$\mu(X_1, \ldots, X_k) = \sigma\left(\kappa\left(m\left(\rho^{-1}(\theta^{-1}(X_1)), \ldots, \rho^{-1}(\theta^{-1}(X_k))\right)\right)\right), X_i \in \mathbb{A}^n.$$

Finally, one can apply the operator $\mathcal{E}$ to express this function as a system of polynomials in sparse form. We obtain $m: (\mathbb{A}^n)^k \to \mathbb{A}^n$
$$m(X_1, \ldots, X_k) = \mathcal{E}\left[\sigma\left(\kappa\left(m\left(\rho^{-1}(\theta^{-1}(X_1)), \ldots, \rho^{-1}(\theta^{-1}(X_k))\right)\right)\right)\right].$$

**Corollary:** *The function $m$, when restricted to elements of $\mathcal{G} \leftrightarrow \{\rho, \theta\}$ is a multilinear form $m: \mathcal{G}^k \to \mathcal{H}$.*

**Example 2**: Suppose $\theta, \rho$ are as above, $\sigma$ is just the identity map, $\kappa(t) = \begin{bmatrix} t \\ 0 \\ 0 \end{bmatrix}$, then $\mathcal{H}$ corresponds the line embedded in space along the first coordinate and the group law is just ordinary addition on that line and

$$\mu\left(\begin{bmatrix}x_1\\x_2\\x_3\end{bmatrix},\begin{bmatrix}y_1\\y_2\\y_3\end{bmatrix},\begin{bmatrix}z_1\\z_2\\z_3\end{bmatrix}\right)=\rho^{-1}\left(\theta^{-1}\left(\begin{bmatrix}x_1\\x_2\\x_3\end{bmatrix}\right)\right)\rho^{-1}\left(\theta^{-1}\left(\begin{bmatrix}y_1\\y_2\\y_3\end{bmatrix}\right)\right)\rho^{-1}\left(\theta^{-1}\left(\begin{bmatrix}z_1\\z_2\\z_3\end{bmatrix}\right)\right).$$ If $v=\theta^{-1}, U=v(X), V=v(Y), W=v(Z), X=\begin{bmatrix}x_1\\x_2\\x_3\end{bmatrix}, Y=\begin{bmatrix}y_1\\y_2\\y_3\end{bmatrix}$ etc., then $m(X,Y,Z)=u_1v_1w_1=$

$2 x_2 y_2 z_1 z_2 + x_2 y_2 z_3 z_1 + 1800 x_2 y_2 z_3 z_2 + 4 x_2 y_3 z_1 z_2 + 2 x_2 y_3 z_3 z_1 + 1799 x_2 y_3 z_3 z_2 + 1799 x_2 y_1^2 z_1 z_2$
$+ 1800 x_2 y_1^2 z_3 z_1 + x_2 y_1^2 z_3 z_2 + 1799 x_2 y_2^2 z_1 z_2 + 1800 x_2 y_2^2 z_3 z_1 + x_2 y_2^2 z_3 z_2 + 1793 x_3 y_1 z_1 z_2$
$+ 1797 x_3 y_1 z_3 z_1 + 4 x_3 y_1 z_3 z_2 + 4 x_3 y_2 z_1 z_2 + 2 x_3 y_2 z_3 z_1 + 1799 x_3 y_2 z_3 z_2 + 8 x_1 y_1 z_1 z_2$
$+ 4 x_1 y_1 z_3 z_1 + 1797 x_1 y_1 z_3 z_2 + 1797 x_1 y_2 z_1 z_2 + 1799 x_1 y_2 z_3 z_1 + 2 x_1 y_2 z_3 z_2 + 8 x_3 y_3 z_1 z_2$
$+ 4 x_3 y_3 z_3 z_1 + 1797 x_3 y_3 z_3 z_2 + 1797 x_3 y_1^2 z_1 z_2 + 1799 x_3 y_1^2 z_3 z_1 + 2 x_3 y_1^2 z_3 z_2 + 1797 x_3 y_2^2 z_1 z_2$
$+ 1799 x_3 y_2^2 z_3 z_1 + 2 x_3 y_2^2 z_3 z_2 + 4 x_1^2 y_1 z_1 z_2 + 2 x_1^2 y_1 z_3 z_1 + 1799 x_1^2 y_1 z_3 z_2 + 1799 x_1^2 y_2 z_1 z_2$
$+ 1800 x_1^2 y_2 z_3 z_1 + x_1^2 y_2 z_3 z_2 + 1797 x_1^2 y_3 z_1 z_2 + 1799 x_1^2 y_3 z_3 z_1 + 2 x_1^2 y_3 z_3 z_2 + 2 x_1^2 y_1^2 z_1 z_2 + x_1^2$
$y_1^2 z_3 z_1 + 1800 x_1^2 y_1^2 z_3 z_2 + 2 x_1^2 y_2^2 z_1 z_2 + x_1^2 y_2^2 z_3 z_1 + 1800 x_1^2 y_2^2 z_3 z_2 + 4 x_2^2 y_1 z_1 z_2 + 2 x_2^2 y_1 z_3 z_1$
$+ 1799 x_2^2 y_1 z_3 z_2 + 1799 x_2^2 y_2 z_1 z_2 + 1800 x_2^2 y_2 z_3 z_1 + x_2^2 y_2 z_3 z_2 + 1797 x_2^2 y_3 z_1 z_2 + 1799 x_2^2 y_3 z_3 z_1$
$+ 2 x_2^2 y_3 z_3 z_2 + 2 x_2^2 y_1^2 z_1 z_2 + x_2^2 y_1^2 z_3 z_1 + 1800 x_2^2 y_1^2 z_3 z_2 + 2 x_2^2 y_2^2 z_1 z_2 + x_2^2 y_2^2 z_3 z_1 + 1800 x_2^2$
$y_2^2 z_3 z_2 + 1793 x_1 y_3 z_1 z_2 + 1797 x_1 y_3 z_3 z_1 + 4 x_1 y_3 z_3 z_2 + 4 x_1 y_1^2 z_1 z_2 + 2 x_1 y_1^2 z_3 z_1 + 1799 x_1$
$y_1^2 z_3 z_2 + 4 x_1 y_2^2 z_1 z_2 + 2 x_1 y_2^2 z_3 z_1 + 1799 x_1 y_2^2 z_3 z_2 + 1797 x_2 y_1 z_1 z_2 + 1799 x_2 y_1 z_3 z_1$
$+ 2 x_2 y_1 z_3 z_2 + 1799 x_1 x_3 y_2 z_1 + 2 x_1 x_3 y_2 z_3 + 1800 x_1 x_3 y_2 z_1^2 + 1800 x_1 x_3 y_2 z_2^2 + 2 x_2 x_3 y_2 z_1$
$+ 1799 x_2 x_3 y_2 z_3 + x_2 x_3 y_2 z_1^2 + x_2 x_3 y_2 z_2^2 + 4 x_1 y_1 y_3 z_1 + 1797 x_1 y_1 y_3 z_3 + 2 x_1 y_1 y_3 z_1^2$
$+ 2 x_1 y_1 y_3 z_2^2 + 1797 x_2 y_1 y_2 z_1 + 4 x_2 y_1 y_2 z_3 + 1799 x_2 y_1 y_2 z_1^2 + 1799 x_2 y_1 y_2 z_2^2 + 1797 x_1 y_2 y_3 z_1$
$+ 4 x_1 y_2 y_3 z_3 + 1799 x_1 y_2 y_3 z_1^2 + 1799 x_1 y_2 y_3 z_2^2 + 2 x_1 x_3 y_2^2 z_1 + 1799 x_1 x_3 y_2^2 z_3 + x_1 x_3 y_2^2 z_1^2$
$+ x_1 x_3 y_2^2 z_2^2 + 4 x_1 x_3 y_1 z_1 + 1797 x_1 x_3 y_1 z_3 + 2 x_1 x_3 y_1 z_1^2 + 2 x_1 x_3 y_1 z_2^2 + 1799 x_1 x_3$
$y_1^2 z_3 + x_1 x_3 y_1^2 z_1^2 + x_1 x_3 y_1^2 z_2^2 + 1797 x_1 x_3 y_3 z_1 + 4 x_1 x_3 y_3 z_3 + 1799 x_1 x_3 y_3 z_1^2 + 1799 x_1 x_3 y_3 z_2^2$
$+ 1799 x_2 y_1 y_3 z_1 + 2 x_2 y_1 y_3 z_3 + 1800 x_2 y_1 y_3 z_1^2 + 1800 x_2 y_1 y_3 z_2^2 + 1799 x_1^2 y_2 y_3 z_1 + 2 x_1^2 y_2 y_3 z_3$
$+ 1800 x_1^2 y_2 y_3 z_1^2 + 1800 x_1^2 y_2 y_3 z_2^2 + 2 x_2^2 y_1 y_3 z_1 + 1799 x_2^2 y_1 y_3 z_3 + x_2^2 y_1 y_3 z_1^2 + x_2^2 y_1 y_3 z_2^2$
$+ 1793 x_1 x_2 y_3 z_1 + 8 x_1 x_2 y_3 z_3 + 1797 x_1 x_2 y_3 z_1^2 + 1797 x_1 x_2 y_3 z_2^2 + 1797 x_2 x_3 y_1 z_1 + 4 x_2 x_3 y_1 z_3$
$+ 1799 x_2 x_3 y_1 z_1^2 + 1799 x_2 x_3 y_1 z_2^2 + 2 x_1^2 y_1 y_3 z_1 + 1799 x_1^2 y_1 y_3 z_3 + x_1^2 y_1 y_3 z_1^2 + x_1^2 y_1 y_3 z_2^2$
$+ 1797 x_3 y_1 y_3 z_1 + 4 x_3 y_1 y_3 z_3 + 1799 x_3 y_1 y_3 z_1^2 + 1799 x_3 y_1 y_3 z_2^2 + 1799 x_2^2 y_2 y_3 z_1 + 2 x_2^2 y_2 y_3 z_3$
$+ 1800 x_2^2 y_2 y_3 z_1^2 + 1800 x_2^2 y_2 y_3 z_2^2 + 4 x_1 x_2 y_2^2 z_1 + 1797 x_1 x_2 y_2^2 z_3 + 2 x_1 x_2 y_2^2 z_1^2 + 2 x_1 x_2 y_2^2 z_2^2$
$+ 4 x_2^2 y_1 y_2 z_1 + 1797 x_2^2 y_1 y_2 z_3 + 2 x_2^2 y_1 y_2 z_1^2 + 2 x_2^2 y_1 y_2 z_2^2 + 1799 x_2 x_3 y_1^2 z_1 + 2 x_2 x_3 y_1^2 z_3$
$+ 1800 x_2 x_3 y_1^2 z_1^2$

$+ 1800\, x_2 x_3 y_1^2 z_2^2 + 1799\, x_2 x_3 y_2^2 z_1 + 2\, x_2 x_3 y_2^2 z_3 + 1800\, x_2 x_3 y_2^2 z_1^2 + 1800\, x_2 x_3 y_2^2 z_2^2 + 4\, x_3 y_2 y_3 z_1$
$+ 1797\, x_3 y_2 y_3 z_3 + 2\, x_3 y_2 y_3 z_1^2 + 2\, x_3 y_2 y_3 z_2^2 + 8\, x_1 x_2 y_1 z_1 + 1793\, x_1 x_2 y_1 z_3 + 4\, x_1 x_2 y_1 z_1^2$
$+ 4\, x_1 x_2 y_1 z_2^2 + 4\, x_2 x_3 y_3 z_1 + 1797\, x_2 x_3 y_3 z_3 + 2\, x_2 x_3 y_3 z_1^2 + 2\, x_2 x_3 y_3 z_2^2 + 1797\, x_1 x_2 y_2 z_1$
$+ 4\, x_1 x_2 y_2 z_3 + 1799\, x_1 x_2 y_2 z_1^2 + 1799\, x_1 x_2 y_2 z_2^2 + 1793\, x_3 y_1 y_2 z_1 + 8\, x_3 y_1 y_2 z_3 + 1797\, x_3 y_1 y_2 z_1^2$
$+ 1797\, x_3 y_1 y_2 z_2^2 + 4\, x_1 x_2 y_1^2 z_1 + 1797\, x_1 x_2 y_1^2 z_3 + 2\, x_1 x_2 y_1^2 z_1^2 + 2\, x_1 x_2 y_1^2 z_2^2 + 4\, x_1^2 y_1 y_2 z_1$
$+ 1797\, x_1^2 y_1 y_2 z_3 + 2\, x_1^2 y_1 y_2 z_1^2 + 2\, x_1^2 y_1 y_2 z_2^2 + 2\, x_2 y_2 y_3 z_1 + 1799\, x_2 y_2 y_3 z_3 + x_2 y_2 y_3 z_1^2$
$+ 8\, x_1 x_2 y_1 y_2 z_1 z_2 + 4\, x_1 x_2 y_1 y_2 z_3 z_1 + 1797\, x_1 x_2 y_1 y_2 z_3 z_2 + 4\, x_1 x_2 y_1 y_3 z_1 z_2 + 2\, x_1 x_2 y_1 y_3 z_3 z_1$
$+ 1799\, x_1 x_2 y_1 y_3 z_3 z_2 + 1797\, x_1 x_2 y_2 y_3 z_1 z_2 + 1799\, x_1 x_2 y_2 y_3 z_3 z_1 + 2\, x_1 x_2 y_2 y_3 z_3 z_2$
$+ 4\, x_1 x_3 y_1 y_2 z_1 z_2 + 2\, x_1 x_3 y_1 y_2 z_3 z_1 + 1799\, x_1 x_3 y_1 y_2 z_3 z_2 + 2\, x_1 x_3 y_1 y_3 z_1 z_2 + x_1 x_3 y_1 y_3 z_3 z_1$
$+ 1800\, x_1 x_3 y_1 y_3 z_3 z_2 + 1799\, x_1 x_3 y_2 y_3 z_1 z_2 + 1800\, x_1 x_3 y_2 y_3 z_3 z_1 + x_1 x_3 y_2 y_3 z_3 z_2$
$+ 1797\, x_2 x_3 y_1 y_2 z_1 z_2 + 1799\, x_2 x_3 y_1 y_2 z_3 z_1 + 2\, x_2 x_3 y_1 y_2 z_3 z_2 + 1799\, x_2 x_3 y_1 y_3 z_1 z_2$
$+ 1800\, x_2 x_3 y_1 y_3 z_3 z_1 + x_2 x_3 y_1 y_3 z_3 z_2 + 2\, x_2 x_3 y_2 y_3 z_1 z_2 + x_2 x_3 y_2 y_3 z_3 z_1 + 1800\, x_2 x_3 y_2 y_3 z_3 z_2$
$+ 1799\, x_2 y_3 z_2^2 + 1799\, x_2 y_1^2 z_3 + x_2 y_1^2 z_2^2 + 4\, x_2 x_3 y_1 z_1 z_2 + 1797\, x_1 x_2 y_3 z_3 z_2 + 4\, x_1 x_2 y_3 z_3 z_1 +$
$x_2^2 y_1 y_3 z_3 z_2 + 8\, x_1 x_2 y_3 z_1 z_2 + 2\, x_1 y_2 y_3 z_3 z_1 + 1799\, x_1 y_2 y_3 z_3 z_2 + 1799\, x_1 x_3 y_2^2 z_1 z_2 + 1800\, x_1 x_3$
$y_2^2 z_3 z_1 + x_1 x_3 y_2^2 z_3 z_2 + 1797\, x_1 x_3 y_1 z_1 z_2 + 1799\, x_1 x_3 y_1 z_3 z_1 + 2\, x_1 x_3 y_1 z_3 z_2 + 1799\, x_1 x_3 y_1^2 z_1 z_2$
$+ 1800\, x_1 x_3 y_1^2 z_3 z_1 + x_1 x_3 y_1^2 z_3 z_2 + 4\, x_1 x_3 y_3 z_1 z_2 + 2\, x_1 x_3 y_3 z_3 z_1 + 1799\, x_1 x_3 y_3 z_3 z_2$
$+ 2\, x_2 y_1 y_3 z_1 z_2 + x_2 y_1 y_3 z_3 z_1 + 1800\, x_2 y_1 y_3 z_3 z_2 + 2\, x_1^2 y_2 y_3 z_1 z_2 + x_1^2 y_2 y_3 z_3 z_1 + 1800$
$x_1^2 y_2 y_3 z_3 z_2 + 1799\, x_2^2 y_1 y_3 z_1 z_2 + 1800\, x_2^2 y_1 y_3 z_3 z_1 + 2\, x_1 x_3 y_2 z_1 z_2 + x_1 x_3 y_2 z_3 z_1$
$+ 1800\, x_1 x_3 y_2 z_3 z_2 + 1799\, x_2 x_3 y_2 z_1 z_2 + 1800\, x_2 x_3 y_2 z_3 z_1 + x_2 x_3 y_2 z_3 z_2 + 1797\, x_1 y_1 y_3 z_1 z_2$
$+ 1799\, x_1 y_1 y_3 z_3 z_1 + 2\, x_1 y_1 y_3 z_3 z_2 + 4\, x_2 y_1 y_2 z_1 z_2 + 2\, x_2 y_1 y_2 z_3 z_1 + 1799\, x_2 y_1 y_2 z_3 z_2$
$+ 4\, x_1 y_2 y_3 z_1 z_2 + 4\, x_2 y_3 z_3 + 4\, x_3 y_2 z_3 + 1800\, x_2 y_2 z_2^2 + 1800\, x_2 y_2 z_1^2 + 2\, x_2 y_2^2 z_1 + 4\, x_3 y_1 z_1^2$
$+ 1799\, x_2 y_2^2 z_3 + 1799\, x_3 y_2 z_2^2 + 1793\, x_1 y_1 z_1 + 8\, x_1 y_1 z_3 + 1797\, x_1 y_1 z_1^2 + 1797\, x_1 y_1 z_2^2 + 4\, x_1 y_2 z_1$
$+ 1797\, x_1 y_2 z_3 + 2\, x_1 y_2 z_1^2 + 2\, x_1 y_2 z_2^2 + 1793\, x_3 y_3 z_1 + 8\, x_3 y_3 z_3 + 1797\, x_3 y_3 z_1^2 + 1797\, x_3 y_3 z_2^2$
$+ 4\, x_3 y_1^2 z_1 + 1797\, x_3 y_1^2 z_3 + 2\, x_3 y_1^2 z_1^2 + 2\, x_3 y_1^2 z_2^2 + 4\, x_3 y_2^2 z_1 + 1797\, x_3 y_2^2 z_3 + 2\, x_3 y_2^2 z_1^2 + 2\, x_3 y_2^2 z_2^2$
$+ 1797\, x_1^2 y_1 z_1 + 4\, x_1^2 y_1 z_3 + 1799\, x_1^2 y_1 z_1^2 + 1799\, x_1^2 y_1 z_2^2 + 2\, x_1^2 y_2 z_1 + 1799\, x_1^2 y_2 z_3 + x_1^2 y_2 z_1^2 +$
$x_1^2 y_2 z_2^2 + 4\, x_1^2 y_3 z_1 + 1797\, x_1^2 y_3 z_3 + 2\, x_1^2 y_3 z_1^2 + 2\, x_1^2 y_3 z_2^2 + 1799\, x_1^2 y_1^2 z_1 + 2\, x_1^2 y_1^2 z_3 + 1800\, x_1^2 y_1^2 z_1^2$
$+ 1800\, x_1^2 y_1^2 z_2^2 + 1799\, x_1^2 y_2^2 z_1 + 2\, x_1^2 y_2^2 z_3 + 1800\, x_1^2 y_2^2 z_1^2 + 1800\, x_1^2 y_2^2 z_2^2 + 1797\, x_2^2 y_1 z_1 + 4\, x_2^2 y_1 z_3$
$+ 1799\, x_2^2 y_1 z_1^2 + 1799\, x_2^2 y_1 z_2^2 + 2\, x_2^2 y_2 z_1 + 1799\, x_2^2 y_2 z_3 + x_2^2 y_2 z_1^2 + x_2^2 y_2 z_2^2 + 4\, x_2^2 y_3 z_1 + 1797$
$x_2^2 y_3 z_3 + 2\, x_2^2 y_3 z_1^2 + 2\, x_2^2 y_3 z_2^2 + 1799\, x_2^2 y_1^2 z_1 + 2\, x_2^2 y_1^2 z_3 + 1800\, x_2^2 y_1^2 z_1^2 + 1800\, x_2^2 y_1^2 z_2^2 + 1799\, x_2^2$
$y_2^2 z_1 + 2\, x_2^2 y_2^2 z_3 + 1800\, x_2^2 y_2^2 z_1^2 + 1800\, x_2^2 y_2^2 z_2^2 + 8\, x_1 y_3 z_1 + 1793\, x_1 y_3 z_3 + 4\, x_1 y_3 z_1^2 + 4\, x_1 y_3 z_2^2$
$+ 1797\, x_1 y_1^2 z_1 + 4\, x_1 y_1^2 z_3 + 1799\, x_1 y_1^2 z_1^2 + 1799\, x_1 y_1^2 z_2^2 + 1797\, x_1 y_2^2 z_1 + 4\, x_1 y_2^2 z_3 + 1799\, x_1 y_2^2 z_1^2$
$+ 1799\, x_1 y_2^2 z_2^2 + 4\, x_2 y_1 z_1 + 1797\, x_2 y_1 z_3 + 2\, x_2 y_1 z_1^2 + 2\, x_2 y_1 z_2^2 + 2\, x_2 y_1^2 z_1 + 1799\, x_2 y_3 z_1^2 + x_2$
$y_2^2 z_2^2 + 1799\, x_3 y_2 z_1^2 + x_2 y_2^2 z_1^2 + 2\, x_2 y_2 z_3 + 4\, x_3 y_1 z_2^2 + 1793\, x_3 y_1 z_3 + 8\, x_3 y_1 z_1 + 1797\, x_3 y_2 z_1 + x_2$
$y_1^2 z_1^2 + 1797\, x_2 y_3 z_1 + 1799\, x_2 y_2 z_1 + 2\, x_2 x_3 y_1 z_3 z_1 + 1799\, x_2 x_3 y_1 z_3 z_2 + 1799\, x_1^2 y_1 y_3 z_1 z_2 + 1800$
$x_1^2 y_1 y_3 z_3 z_1 + x_1^2 y_1 y_3 z_3 z_2 + 4\, x_3 y_1 y_3 z_1 z_2 + 2\, x_3 y_1 y_3 z_3 z_1 + 1799\, x_3 y_1 y_3 z_3 z_2 + 2\, x_2^2 y_2 y_3 z_1 z_2 +$

$x_2^2 y_2 y_3 z_3 z_1 + 1800 x_2^2 y_2 y_3 z_3 z_2 + 1797 x_1 x_2 y_2^2 z_1 z_2 + 1799 x_1 x_2 y_2^2 z_3 z_1 + 2 x_1 x_2 y_2^2 z_3 z_2 + 1797 x_2^2 y_1 y_2 z_1 z_2 + 1799 x_2^2 y_1 y_2 z_3 z_1 + 2 x_2^2 y_1 y_2 z_3 z_2 + 2 x_2 x_3 y_1^2 z_1 z_2 + x_2 x_3 y_1^2 z_3 z_1 + 1800 x_2 x_3 y_1^2 z_3 z_2 + 2 x_2 x_3 y_2^2 z_1 z_2 + x_2 x_3 y_2^2 z_3 z_1 + 1800 x_2 x_3 y_2^2 z_3 z_2 + 1797 x_3 y_2 y_3 z_1 z_2 + 1799 x_3 y_2 y_3 z_3 z_1 + 2 x_3 y_2 y_3 z_3 z_2 + 1793 x_1 x_2 y_1 z_1 z_2 + 1797 x_1 x_2 y_1 z_3 z_1 + 4 x_1 x_2 y_1 z_3 z_2 + 1797 x_2 x_3 y_3 z_1 z_2 + 1799 x_2 x_3 y_3 z_3 z_1 + 2 x_2 x_3 y_3 z_3 z_2 + 4 x_1 x_2 y_2 z_1 z_2 + 2 x_1 x_2 y_2 z_3 z_1 + 1799 x_1 x_2 y_2 z_3 z_2 + 8 x_3 y_1 y_2 z_1 z_2 + 4 x_3 y_1 y_2 z_3 z_1 + 1797 x_3 y_1 y_2 z_3 z_2 + 1797 x_1 x_2 y_1^2 z_1 z_2 + 1799 x_1 x_2 y_1^2 z_3 z_1 + 2 x_1 x_2 y_1^2 z_3 z_2 + 1797 x_1^2 y_1 y_2 z_1 z_2 + 1799 x_1^2 y_1 y_2 z_3 z_1 + 2 x_1^2 y_1 y_2 z_3 z_2 + 1799 x_2 y_2 y_3 z_1 z_2 + 1800 x_2 y_2 y_3 z_3 z_1 + x_2 y_2 y_3 z_3 z_2 + 1793 x_1 y_1 y_2 z_1 z_2 + 1797 x_1 y_1 y_2 z_3 z_1 + 4 x_1 y_1 y_2 z_3 z_2 + 1793 x_1 x_2 y_1 y_2 z_1 + 8 x_1 x_2 y_1 y_2 z_3 + 1797 x_1 x_2 y_1 y_2 z_1^2 + 1797 x_1 x_2 y_1 y_2 z_2^2 + 1797 x_1 x_2 y_1 y_3 z_1 + 4 x_1 x_2 y_1 y_3 z_3 + 1799 x_1 x_2 y_1 y_3 z_1^2 + 1799 x_1 x_2 y_1 y_3 z_2^2 + 4 x_1 x_2 y_2 y_3 z_1 + 1797 x_1 x_2 y_2 y_3 z_3 + 2 x_1 x_2 y_2 y_3 z_1^2 + 2 x_1 x_2 y_2 y_3 z_2^2 + 1797 x_1 x_3 y_1 y_2 z_1 + 4 x_1 x_3 y_1 y_2 z_3 + 1799 x_1 x_3 y_1 y_2 z_1^2 + 1799 x_1 x_3 y_1 y_2 z_2^2 + 1799 x_1 x_3 y_1 y_3 z_1 + 2 x_1 x_3 y_1 y_3 z_3 + 1800 x_1 x_3 y_1 y_3 z_1^2 + 1800 x_1 x_3 y_1 y_3 z_2^2 + 2 x_1 x_3 y_2 y_3 z_1 + 1799 x_1 x_3 y_2 y_3 z_3 + x_1 x_3 y_2 y_3 z_1^2 + x_1 x_3 y_2 y_3 z_2^2 + 4 x_2 x_3 y_1 y_2 z_1 + 1797 x_2 x_3 y_1 y_2 z_3 + 2 x_2 x_3 y_1 y_2 z_1^2 + 2 x_2 x_3 y_1 y_2 z_2^2 + 2 x_2 x_3 y_1 y_3 z_1 + 1799 x_2 x_3 y_1 y_3 z_3 + x_2 x_3 y_1 y_3 z_1^2 + x_2 x_3 y_1 y_3 z_2^2 + 1799 x_2 x_3 y_2 y_3 z_1 + 2 x_2 x_3 y_2 y_3 z_3 + 1800 x_2 x_3 y_2 y_3 z_1^2 + 1800 x_2 x_3 y_2 y_3 z_2^2 + x_2 y_2 y_3 z_2^2 + 8 x_1 y_1 y_2 z_1 + 1793 x_1 y_1 y_2 z_3 + 4 x_1 y_1 y_2 z_1^2 + 4 x_1 y_1 y_2 z_2^2 + x_1 x_3 y_1 y_3 z_2 + 1800 x_1 x_3 y_2 y_3 z_2 + 1799 x_2 x_3 y_1 y_2 z_2 + 1800 x_2 x_3 y_1 y_3 z_2 + x_2 x_3 y_2 y_3 z_2 + 4 x_1 x_2 y_1 y_2 z_2 + 2 x_1 x_2 y_1 y_3 z_2 + 1799 x_1 x_2 y_2 y_3 z_2 + 2 x_1 x_3 y_1 y_2 z_2 + 2 x_3 y_2 z_2 + x_2 y_2 z_2 + 1800 x_2 y_1^2 z_2 + 1800 x_2 y_2^2 z_2 + 1797 x_3 y_1 z_2 + 2 x_2 y_3 z_2 + 4 x_1 y_1 z_2 + 1799 x_1 y_2 z_2 + 4 x_3 y_3 z_2 + 1799 x_3 y_1^2 z_2 + 1799 x_3 y_2^2 z_2 + 2 x_1^2 y_1 z_2 + 1800 x_1^2 y_2 z_2 + 1799 x_1^2 y_3 z_2 + x_1^2 y_1^2 z_2 + x_1^2 y_2^2 z_2 + 2 x_2^2 y_1 z_2 + 1800 x_2^2 y_2 z_2 + 1799 x_2^2 y_3 z_2 + x_2^2 y_1^2 z_2 + x_2^2 y_2^2 z_2 + 1797 x_1 y_3 z_2 + 2 x_1 y_1^2 z_2 + 2 x_1 y_2^2 z_2 + 1799 x_2 y_1 z_2 + x_1 x_3 y_2 z_2 + 1800 x_2 x_3 y_2 z_2 + 1799 x_1 y_1 y_3 z_2 + 2 x_2 y_1 y_2 z_2 + 2 x_1 y_2 y_3 z_2 + 1800 x_1 x_3 y_2^2 z_2 + 1799 x_1 x_3 y_1 z_2 + 1800 x_1 x_3 y_1^2 z_2 + 2 x_1 x_3 y_3 z_2 + x_2 y_1 y_3 z_2 + x_1^2 y_2 y_3 z_2 + 1800 x_2^2 y_1 y_3 z_2 + 4 x_1 x_2 y_3 z_2 + 2 x_2 x_3 y_1 z_2 + 1800 x_1^2 y_1 y_3 z_2 + 2 x_3 y_1 y_3 z_2 + x_2^2 y_2 y_3 z_2 + 1799 x_1 x_2 y_2^2 z_2 + 1799 x_2^2 y_1 y_2 z_2 + x_2 x_3 y_1^2 z_2 + x_2 x_3 y_2^2 z_2 + 1799 x_3 y_2 y_3 z_2 + 1797 x_1 x_2 y_1 z_2 + 1799 x_2 x_3 y_3 z_2 + 2 x_1 x_2 y_2 z_2 + 4 x_3 y_1 y_2 z_2 + 1799 x_1 x_2 y_1^2 z_2 + 1799 x_1^2 y_1 y_2 z_2 + 1800 x_2 y_2 y_3 z_2 + 1797 x_1 y_1 y_2 z_2$

One practical way to construct these groups is to use *tame* automorphisms [M1], i.e. $\theta: \mathbb{A}^n \xrightarrow{\sim} \mathbb{A}^n$ where $\theta = \lambda_1 \circ \tau \circ \lambda_2$ where $\lambda_1, \lambda_2$ are linear functions and $\tau(x_1, \ldots, x_n) = (x_1 + \tau_1, x_2 + \tau_2, \ldots, x_n + \tau_n)$, is a triangular transformation with each polynomial $\tau_i$ a polynomial in the variables $\{x_j, j < i\}$. Denote the set of such triangular morphisms by $t_{p,d,n}$ and the corresponding set of tame transformation as $\mathcal{T}_{p,d,n}$. Let $\mathcal{I}_n$ denote the set of all linear immersions $\rho: \mathbb{A}^1 \to \mathbb{A}^n$. Define $\mathfrak{T}_{p,d,n}$ as the set of group structures that can be constructed with the method defined above using tame transformations in $\mathcal{T}_{p,d,n}$ and using a linear immersions in $\mathcal{I}_n$. Let $\mathfrak{T}_{p,n} = \cup_d \mathfrak{T}_{p,d,n}$, $\mathfrak{T}_p = \cup_n \mathfrak{T}_{p,n}$. We have $\mathfrak{T}_{p,d,n} \subset \mathfrak{G}_{p,d,n}, \mathfrak{T}_{p,n} \subset \mathfrak{G}_{p,n}, \mathfrak{T}_p \subset \mathfrak{G}_p$.

We examine the size of the key space of these groups with the following

**Proposition**: $|\mathfrak{T}_{p,d,n}| \geq p^{\left(2n + \prod_{1 \leq i \leq n} \binom{d+i}{d}\right)}$.

Proof: Since a triangular transformation is of the form $\tau(x_1, \ldots, x_n) = (x_1 + \tau_1, x_2 + \tau_2, \ldots, x_n + \tau_n)$ with each $\tau_i$ a polynomial of total degree $\leq d$ in the variables $\{x_j, j < i\}$, $t_{p,d,n} \cong \prod_{i < n} P_{i,d}$, where $P_{i,d}$ denotes polynomials in $i$ variables of total degree $\leq d$. Therefore $t_{p,d,n} = p^{\prod_{1 \leq i \leq n} \binom{d+i}{d}}$. There are $p^{2n}$ distinct linear immersions. The set of possible groups include all those corresponding to any $\tau \in t_{p,d,n}$ and any linear immersion and such groups are distinct from each other. The proposition follows.

**Corollary**: *If $\mathcal{G} \in \mathfrak{G}_{p,d,n}$ is chosen with a uniform distribution then its entropy satisfies $H(\mathcal{G}) \geq$* $\log(p) \, p^{\prod_{1 \leq i \leq n} \binom{d+i}{d}}$.

This shows that the information available from a polynomial amount of input-output data to determine a group is far exceeded by the uncertainly of possible group laws as the parameter $n$ grows.

These types of group presentations can be generalized by attaching group structure to any rational curve which is isomorphic to an embedding of $\mathbb{A}^1$ into $\mathbb{A}^n$.

**Proposition:** *Given any curve $C$ in $\mathbb{A}^n$ over $\mathbb{F} = GF(p)$ which is isomorphic to $\mathbb{A}^1$ and two points $P_0, P_1 \in C(\mathbb{F})$, there is a unique group structure on $C(\mathbb{F})$ making it isomorphic to the group structure on $\mathbb{A}^1(\mathbb{F})$, carrying $P_0$ to $0$ and $P_1$ to $1$.*

Proof: A group structure on $C$ is derived from an isomorphism $\rho: \mathbb{A}^1 \xrightarrow{\sim} C$. Two distinct group structures would give rise to an isomorphism $\mathbb{A}^1 \xrightarrow{\sim} \mathbb{A}^1$. The only such isomorphisms correspond to affine maps, which are determined by the correspondence between two pairs of points.

This allows for the extension of the encryption algorithm beyond those requiring automorphisms, one could simply use the group structure on any rational curve.

An interesting question is whether the group structure on any such rational curve corresponds to some group in $\mathfrak{G}$.

We next take up the question of the security of this encryption algorithm in low dimensional situations. We do this by examining some of the implications of existence of a discrete logarithm.

**Discrete Logarithm Definitions:**
1. Given a group $\mathcal{G} \in \mathfrak{G}_p$ with group law $\mathcal{g}$ and generator $P_0$ a discrete logarithm of type 1 is an algebraic circuit $\mathbb{e}: \mathbb{F}^n \to \mathbb{F}^1$ such that $\mathbb{e}(Q) = e$ if $Q = P_0^e$ in $\mathcal{G}$.
2. Given a group $\mathcal{G} \in \mathfrak{G}_p$ with group law $\mathcal{g}$ with corresponding automorphism $\theta: \mathbb{A}^n \xrightarrow{\sim} \mathbb{A}^n$ and affine line $C$ with parameterization $\sigma$, a discrete logarithm of type 2 and degree d is a polynomial $\delta \in \mathbb{F}[x_1, \cdots x_n]$ of total degree d that satisfies $\delta \circ \theta \circ \sigma(t) = t$ for $t \in \mathbb{A}^1(\mathbb{F})$.
3. Given a group $\mathcal{G} \in \mathfrak{G}_p$ with group law $\mathcal{g}$ with corresponding automorphism $\theta: \mathbb{A}^n \xrightarrow{\sim} \mathbb{A}^n$ and affine line $C$ with parameterization $\sigma$, a discrete logarithm of type 3 and degree d is a polynomial $\delta \in \mathbb{F}[x_1, \cdots x_n]$ of total degree d which when restricted to $D = \theta(C)$, induces an algebraic inverse of $\rho = \theta \circ \sigma$.
4. Given a group structure $\mathcal{G}$ on a rational curve $D \subset \mathbb{A}^n$ corresponding to an embedding of $\mathbb{A}^1$ into $\mathbb{A}^n$, a discrete logarithm of type 4 and degree d is a polynomial $\delta \in \mathbb{F}[x_1, \cdots x_n]$ of total degree d which, when restricted to D, gives an isomorphism $\delta|_D \xrightarrow{\sim} \mathbb{A}^1$ carrying the group structure $\mathcal{G}$ on $D(\mathbb{F})$ to the additive group structure on $\mathbb{A}^1(\mathbb{F})$.

Note that these definitions of discrete logarithm are increasingly restrictive. Note also that discrete logarithm of types 3 or 4 is essentially a uniformizing parameter [SH1] for the underlying rational curve. We relate the existence of a discrete logarithm to the problem of inverting systems of equations via:

**Inversion Problem:** Given an automorphism $\theta: \mathbb{A}^n \xrightarrow{\sim} \mathbb{A}^n$ and $Q \in \mathbb{A}^n(\mathbb{F})$, find $P \in \mathbb{A}^n(\mathbb{F})$ with $\theta(P) = Q$.

**Inversion Algorithm:** Start with an automorphism $\theta: \mathbb{A}^n \xrightarrow{\sim} \mathbb{A}^n$ and $Q \in \mathbb{A}^n(\mathbb{F})$, Choose points $P_0, P_1 \in \mathbb{F}^n$ at random and let $Q_i = \theta(P_i)$. Let $H$ be the plane containing $Q_1, Q_2$ and $Q$ and let $Y$ be the conic on $H$ that passes through $Q_0, Q_1$ and $Q$ and has a tangency to the line at infinity of $H$. Suppose that $\sigma: \mathbb{A}^1 \to \mathbb{A}^n$ parametrizes $Y$ and that $\sigma(0) = Q_0, \sigma(1) = Q_1$ and $\sigma(t) = Q$. Let $\langle f_i(x_1, \ldots, x_n)\rangle_{i=1,\ldots,n-1}$ be the ideal defining $Y$ in $\mathbb{A}^n$, where each polynomial $f_i$ is linear or quadratic. Then the pre-image $X = \theta^*(Y)$ is a rational curve which contains the points $P_0, P_1$ as well as the pre-image $P = \theta^{-1}(Q)$. $X$ is defined by the ideal $\langle f_i(\theta_1, \ldots, \theta_n)\rangle_{i=1,\ldots,n-1}$. Suppose one had a polynomial $\delta: \mathbb{A}^n \to \mathbb{A}^1$ which induced an isomorphism $\delta: X \xrightarrow{\sim} \mathbb{A}^1$ so that $\delta(P_0) = 0, \delta(P_1) = 1$. Then one would have $\delta(P) = t$. Using the lex ordering on the variables $x_1, \ldots, x_n$ it is clear that a Groebner basis for the ideal $I = \langle f_1(\theta_1, \ldots, \theta_n), \ldots, f_{n-1}(\theta_1, \ldots, \theta_n), \delta(x_1, \ldots, x_n) - t\rangle$ is of the form $\langle x_1 - a_1, \ldots, x_n - a_n\rangle$ for some $a_i \in \mathbb{F}$. Then $P = (a_1, \ldots, a_n)$.

The algorithm yields the following:

**Theorem:** *Suppose an adversary can produce a discrete logarithm of type 4 for any group $\mathcal{G}$ corresponding to a rational curve in $\mathbb{A}^n$. Then that adversary can be used to invert any automorphism $\theta \in Aut(\mathbb{A}^n)$.*

The computational challenge in implementing this algorithm is obtaining a Groebner basis for the zero-dimensional ideal $I$. Let $J$ denote the homogenized version of $I$.

**Corollary**: *If every component of an automorphism $\theta \in Aut(\mathbb{A}^n)$ has total degree $\leq d$ and an adversary as in the preceding exists who can produce a discrete logarithm of type 4 and degree d then the ideal $J$ in the inversion algorithm is 2d-regular and zero dimensional.*

Proof: Since every polynomial $f_i$ is linear or quadratic, the polynomials $f_i(\theta_1, \ldots, \theta_n)$ have degree at most $2d$. If the discrete logarithm is of type 4 and degree $d$, each member of $\{f_1(\theta_1, \ldots, \theta_n), \ldots, f_{n-1}(\theta_1, \ldots, \theta_n), \delta - t\}$ has degree at most $2d$. By construction, the ideal $J$ corresponds to the closed point $P$. Choose linear combinations $y_0, \ldots, y_n$ of $x_0, \ldots, x_n$ so that $y_0 = x_0$ and $P \notin \{y_i = 0\}, i > 0$. If $g$ is homogenous of degree $2d$ and $gy_i \in J$ then $g(P)y_i(P) = 0$ so $g(P) = 0$ so $g \in J$. In the affine set $x_0 \neq 0$, so $y_0$ corresponds to the constant function 1 in affine coordinates, so for every polynomial $g$ of degree $2d$ in those coordinates $g \in I + \sum_i y_i \mathbb{F}[x]$, so the corresponding statement is true in homogenous coordinates. This shows that $J$ meets the Bayer-Stillman criteria for *2d*-regularity [BS2].

Regularity is a good proxy for ease of computing a Groebner basis [BM1]. We conclude that an adversary that can produce discrete logarithms of type 4 and low degree can easily invert automorphisms of affine spaces.

We next take up the complexity of this algorithm:

**Proposition:** *If $\mathcal{G} \in \mathfrak{T}_{p,d,n}$ then $\deg(\mathcal{g}) \leq d^n$.*

Proof: Since $\mathcal{G} \in \mathfrak{T}_{p,d,n}$, the group law is given by $\mathcal{g}(x,y) = \mathcal{E}\left[\theta\left(\rho\left(\rho^{-1}(\theta^{-1}(x)) + \rho^{-1}(\theta^{-1}(y))\right)\right)\right]$ with $\theta = \lambda_1 \circ \tau \circ \lambda_2$. Here $\lambda_1, \lambda_2$ and linear and $\tau(x_1, \ldots, x_n) = (x_1 + \tau_1, x_2 + \tau_2, \ldots, x_n + \tau_n)$ with each $\tau_i$ a polynomial in the variables $\{x_j, j < i\}$ and total degree $\leq d$. Then $\tau^{-1}(x_1, \ldots, x_n) = (x_1 - \tau_1, x_2 - \tau_2(x_1 - \tau_1), \ldots, x_n - \tau_n(x_1 - \tau_1, x_2 - \tau_2(x_1 - \tau_1), \ldots))$. From this, it is clear that $\deg(\tau^{-1}) \leq d^{n-1}$ and so $\deg(\theta^{-1}) \leq d^{n-1}$. Since $\rho$ and $\rho^{-1}$ are linear the proposition follows.

There are several possible avenues to reduce the exponential explosion in group laws using large dimensions. One is the use of sparse linear transformations and triangular transformations as components of the automorphism.

In summary this paper introduces a class of cyclic groups based on rational curves and non-linear automorphisms of affine space. It discussed how these groups can be used to create cryptographic multilinear maps. It discusses various versions of discrete logarithms for these groups and shows how the existence of some versions discrete logarithms lead to efficient methods of inverting non-linear automorphisms.